# Influences of Divalent Ions in Natural Seawater/River Water on Nanofluidic Osmotic Energy Generation

Fenhong Song,[1] Xuan An,[1,2] Long Ma,[2] Jiakun Zhuang,[2] and Yinghua Qiu [2,3,4,5]*

1. School of Energy and Power Engineering, Northeast Electric Power University, Jilin 132012, China

2. Key Laboratory of High Efficiency and Clean Mechanical Manufacture of Ministry of Education, National Demonstration Center for Experimental Mechanical Engineering Education, School of Mechanical Engineering, Shandong University, Jinan 250061, China

3. Shenzhen Research Institute of Shandong University, Shenzhen 518000, China

4. Suzhou Research Institute, Shandong University, Suzhou 215123, China

5. Key Laboratory of Ocean Energy Utilization and Energy Conservation of Ministry of Education, Dalian 116024, China

*Corresponding author: yinghua.qiu@sdu.edu.cn

**Abstract:**

Besides the dominant NaCl, natural seawater/river water contains trace multivalent ions, which can provide effective screening to surface charges. Here, in both negatively and positively charged nanopores, influences from divalent ions as counterions and coions have been investigated on the performance of osmotic energy conversion (OEC) under natural salt gradients. As counterions, trace $Ca^{2+}$ ions can suppress the electric power and conversion efficiency significantly. The reduced OEC performance is due to the bivalence and low diffusion coefficient of $Ca^{2+}$ ions, instead of the uphill transport of divalent ions discovered in the previous work. Effectively screened charged surfaces by $Ca^{2+}$ ions induce enhanced diffusion of $Cl^-$ ions which simultaneously decreases the net ion penetration and ionic selectivity of the nanopore. While as coions, $Ca^{2+}$ ions have weak effects on the OEC performance. The promotion from charged exterior surfaces on OEC processes for ultra-short nanopores is also studied, which effective region is ~200 nm in width beyond pore boundaries independent of the presence of $Ca^{2+}$ ions. Our results shed light on the physical details of the nanofluidic OEC process under natural seawater/river water conditions, which can provide a useful guide for high-performance osmotic energy harvesting.

**Keywords:** Osmotic energy conversion, Natural salt gradient, Divalent ions, Electric double layers, Nanopores

## 1. Introduction

Osmotic energy[1,2] existing widely between two water bodies under salt gradients, such as seawater and river water, has attracted extensive attention, which can be effectively harvested through the reverse electrodialysis (RED) technique using ion-selective porous membranes.[3-6] The ionic selectivity of nanopores originates from the surface charges on pore walls.[7] Due to the strong electrostatic interaction between surface charges and free ions, counterions are attracted to charged surfaces to form electric double layers (EDLs).[8,9] Inside the nanopore, EDLs regions provide a high-speed passageway for the diffusive transport of ions under salt gradients.[10] During the process of osmotic energy conversion (OEC) the counterions on the high-concentration side acting as the main carriers diffuse through the nanopore to the low-concentration side, and a considerable potential difference is induced across the nanoporous membrane. To achieve high-performance OEC, porous membranes with both high ionic selectivity and large ion permeability are usually required simultaneously.[11,12]

With the rapid development in nanofabrication, various nanoporous membranes have been produced such as polymer membranes,[13] graphene films,[14] and wood membranes,[15] which can be modified chemically to modulate the surface charge properties.[16,17] Single charged nanoporous membranes are usually applied for osmotic energy harvesting, during which the directional transport of anions or cations is used for energy conversion. According to the classical theoretical predictions, the output voltage and power are positively correlated to the diffusion coefficient of counterions.[3,4] Compared to $Na^+$ ions, OEC with $Cl^-$ ions as the main current carriers can produce higher electric power.[18] A positively charged porous membrane can form a membrane pair with a negatively charged membrane, which can be used for OEC with both anions and cations.[14,15] Due to the series connection of both membranes in the circuit, the output membrane potential is the sum of individual ones. By connecting multiple membrane pairs, the output voltage can be regulated linearly to the volt level which can be used to power a calculator or LED lights.[14, 15]

Until now, artificial seawater and river water have been used to supply the salt gradient for most studies on OEC, i.e. NaCl or KCl solutions.[16, 19, 20] However, trace ions with multivalence in natural seawater/river water have been rarely considered. In real situations, besides $Na^+$ and $Cl^-$ ions, seawater/river water contains multiple kinds of trace cations and anions, of which the main ones are divalent $Mg^{2+}$, $Ca^{2+}$, and $SO_4^{2-}$ ions. Divalent cations in seawater can reach up to 50 mM in concentration, which is higher than that of $SO_4^{2-}$ ions ~25 mM.[21, 22] Compared to monovalent ions, due to the much stronger electrostatic interactions between divalent ions and surface charges, a more significant accumulation of divalent cations is induced in the EDLs near charged pore walls, which provide a much better electrostatic screening to surface charges.[8] Excessive aggregation of divalent ions even can induce an inverted surface charge density, i.e. charge inversion.[23-25] Due to that the surface charges determine the ion distributions inside the EDLs, the appearance of divalent ions in natural seawater and river water may have significant influences on OEC processes.

In the experimental work conducted by Vermass et al.[26] trace divalent ions decreased the OEC performance, which they thought was attributed to the uphill transport of divalent ions from the low-concentration side to the high-concentration side. However, their explanation is hard to be adopted during the OEC process, because in their experiments the uphill transport of divalent ions was found when no current density was generated i.e. open-circuit state, and with the current density increasing the uphill transport decreased sharply. Here, the effects of divalent ions on the OEC performance have been investigated with mixed solutions through finite element simulations. Due to the similar diffusion coefficients of both $Ca^{2+}$ and $Mg^{2+}$ ions,[27] $Ca^{2+}$ ions were added to the NaCl solutions as representative divalent ions according to their natural concentration ratio in seawater and river water. Please note that because of the total concentration of divalent ions ~50 mM in seawater and the weak surface charge densities used in our simulations, charge inversion near charged pore walls has not been considered in our simulations,[28-31] which also did not appear in the earlier experimental works with natural seawaters.[17, 32] When the nanopore is negatively charged, $Ca^{2+}$ ions as counterions are attracted to the charged surface more easily and provide a better electrostatic screening to surface charges. Much more $Cl^-$ ions accumulate inside the nanopore which can pass through the nanopore and induce greatly reduced OEC performance. The output power and energy conversion efficiency are suppressed seriously, respectively. This is mainly due to the bivalence of $Ca^{2+}$ ions besides their lower diffusion coefficient. While, for positively charged nanopores, the appearance of $Ca^{2+}$ ions in aqueous solutions as coions has almost no effect on the OEC performance. For the ultra-short nanopores used in our work, charged exterior surfaces on the low-concentration side could play an important role in OEC, which can promote electric power and conversion efficiency simultaneously. Through consideration of the charged ring regions near pore boundaries, the effective charged area on the exterior membrane surface is found at ~200 nm which exhibits no dependence on the addition of $Ca^{2+}$ ions or surface charge polarity.

## 2. Simulation Details

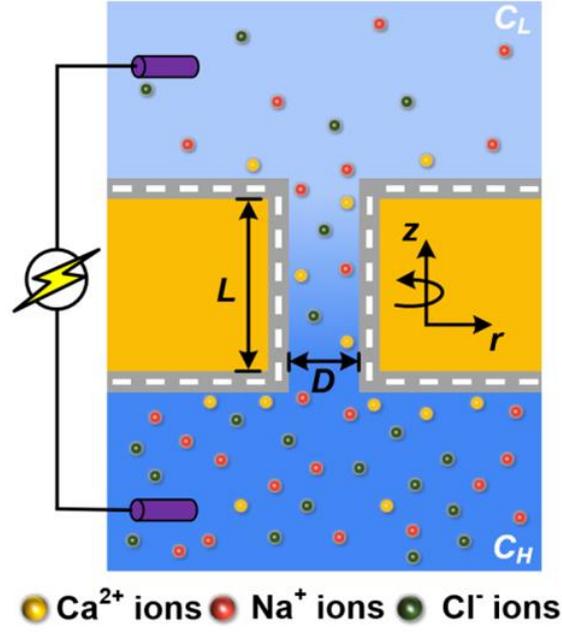

Figure 1. Simulation scheme for nanofluidic osmotic energy conversion under natural concentration gradients with mixed NaCl and CaCl$_2$ solution. Darker and lighter blue regions on both sides of nanopores show the aqueous solutions with a higher ($C_H$) and lower ($C_L$) salt concentration. $L$ and $D$ represent the length and diameter of nanopores in simulations.

Figure 1 shows the scheme of our nanofluidic simulation model. The charged nanopore locates between two reservoirs with a higher ($C_H$) and lower ($C_L$) salt concentration. To comprehensively consider the ionic distributions at the solid-liquid interfaces and the influence of fluid flow on ionic transport through nanopores,[33, 34] coupled Poisson-Nernst-Planck and Navier-Stokes equations were solved through COMSOL Multiphysics with equations 1-4.

$$\varepsilon \nabla^2 \varphi = -\sum_{i=1}^{N} z_i F C_i \quad (1)$$

$$\nabla \cdot \mathbf{J}_i = \nabla \cdot \left( C_i \mathbf{u} - D_i \nabla C_i - \frac{F z_i C_i D_i}{RT} \nabla \varphi \right) = 0 \quad (2)$$

$$\mu \nabla^2 \mathbf{u} - \nabla p - \sum_{i=1}^{N} (z_i F C_i) \nabla \varphi = 0 \quad (3)$$

$$\nabla \cdot \mathbf{u} = 0 \tag{4}$$

where $\varepsilon$, $\varphi$, $N$, $F$, $R$, $T$, $\mathbf{u}$, and $p$ are the permittivity, electrical potential, number of ion species, Faraday's constant, gas constant, temperature, fluid velocity, and pressure, respectively. $z_i$, $C_i$, $\mathbf{J}_i$, and $D_i$, are the valence, concentration, ionic flux, and diffusion coefficient of ionic species $i$ ($Na^+$, $Ca^{2+}$, and $Cl^-$ ions), respectively.

In simulation systems, the cylindrical nanopore provides a channel to connect both reservoirs. The diameter of the nanopore was set as 10 nm[35-39] and the pore length varied from 1 nm to 1000 nm. For reservoirs, the radius and height were chosen as 5 μm which is large enough to ensure the simulation correctness in the cases with long nanopores.[40] The permittivity and temperature of water were set as 80 and 298 K, respectively. To simulate a natural situation and effectively guide practical applications, 500 mM and 10 mM NaCl solutions were added to reservoirs as $C_{Na\_H}$ and $C_{Na\_L}$, respectively.[41, 42] Trace divalent ions in seawater and river water were considered with $Ca^{2+}$ ions. The salt gradients of $CaCl_2$ concentration in the high- and low-concentration reservoirs were fixed at 50, with the concentration of $CaCl_2$ solution at the high-concentration side ($C_{Ca\_H}$) varying from 10 mM to 50 mM. Both negatively and positively charged nanopores were used to investigate the effects of $Ca^{2+}$ ions on the performance of OEC as counterions and coions. The corresponding surface charge density was set to ±0.08 C/m².[43, 44] Diffusion coefficients of $Na^+$, $Ca^{2+}$, and $Cl^-$ were $1.33 \times 10^{-9}$, $0.79 \times 10^{-9}$, and $2.03 \times 10^{-9}$ m²/s, respectively.[27]

Detailed boundary conditions are listed in Table S1. A similar mesh strategy to our earlier works[10, 45-47] was selected as shown in Figure S1. For charged surfaces, the mesh size of 0.1 nm was used which is small enough to consider the ion distributions in electric double layers. For each simulation, the total number of grid nodes exceeded 1,000,000, which is sufficient to ensure the result's accuracy.[46, 48, 49] By conducting the same study as Cao et al.[50] our simulation model had been verified (Figure S2). Also, in our earlier works, the influences of individual charged surfaces[10] and slippery surfaces[46] had been investigated with similar simulation models under the concentration gradients of NaCl.

During the OEC process, the maximum output electric power ($P_{max}$)[3, 4] is one of the important parameters to evaluate the OEC performance, it can be obtained by equation 5.

$$P_{max} = \frac{1}{4}IV \qquad (5)$$

in which $I$ and $V$ are the steady-state diffusion current and membrane potential across the nanopore. $I$ was obtained by integrating the total ionic flux over the reservoir boundary. For solutions with only monovalent counterions, equation 6 provides the theoretical prediction for the membrane potential ($V$).[50] Due to the presence of $Ca^{2+}$ ions as counterions in mixed solutions, the predicted membrane potential from equation 6 is slightly different from the intercept of current-voltage (IV) curves (Figure S3-S4). In the cases containing negatively charged nanopores with $CaCl_2$, the membrane potential was extracted as the intercept of IV curves. The membrane potential was predicted from equation 6 for the other simulated cases including I. a negatively charged nanopore in NaCl solution, II. a positively charged nanopore in NaCl solutions, and III. a positively charged nanopore in NaCl solutions with $CaCl_2$. Equations 7 and 8 can be used to describe the OEC efficiency ($\eta$),[50] of which equation 7 is applied for the cases with negatively charged nanopores immersed in mixed NaCl and $CaCl_2$ solutions, and equation 8 is used for the other cases.

$$V = \left( t_{Na} - t_{Cl} + \frac{1}{2}t_{Ca} \right) \frac{RT}{F} \ln \delta \qquad (6)$$

$$\eta = \frac{I \cdot V}{\dfrac{RT}{F}\left( |I_{Na}| + \frac{1}{2}|I_{Ca}| + |I_{Cl}| \right) \ln \delta} \qquad (7)$$

$$\eta = \frac{(|I_{Na}|+|I_{Ca}|-|I_{Cl}|) \cdot (|I_{Na}|+\frac{1}{2}|I_{Ca}|-|I_{Cl}|)}{(|I_{Na}|+|I_{Ca}|+|I_{Cl}|) \cdot (|I_{Na}|+\frac{1}{2}|I_{Ca}|+|I_{Cl}|)} \tag{8}$$

where $\delta$ is the ratio of ionic chemical activity of ionic species $i$ (Na$^+$, Ca$^{2+}$, and Cl$^-$ ions) on both high-concentration ($\alpha_{i\_H}$) and low-concentration ($\alpha_{i\_L}$) sides, i.e. $\delta = \alpha_{i\_H}/\alpha_{i\_L}$. Here, $\delta$ was kept at 50. $I_{Na}$, $I_{Ca}$, and $I_{Cl}$ are the ionic currents contributed by Na$^+$ ions, Ca$^{2+}$ ions, and Cl$^-$ ions, respectively. $t_{Na}$, $t_{Ca}$, $t_{Cl}$, and $t$ are the transfer number of Na$^+$ ions, Ca$^{2+}$ ions, Cl$^-$ ions, and counterions,[7] which can be evaluated with equations 9-12. For charged nanopores, $t$ varies from 0 to 1. At $t = 1$, the nanopore exhibits 100% selectivity to counterions.

$$t_{Na} = |I_{Na}|/(|I_{Na}|+|I_{Ca}|+|I_{Cl}|) \tag{9}$$

$$t_{Ca} = |I_{Ca}|/(|I_{Na}|+|I_{Ca}|+|I_{Cl}|) \tag{10}$$

$$t_{Cl} = |I_{Cl}|/(|I_{Na}|+|I_{Ca}|+|I_{Cl}|) \tag{11}$$

$$t = |I_{counterion}|/(|I_{counterion}|+|I_{coion}|) \tag{12}$$

where $I_{counterion}$ and $I_{coion}$ are the current values contributed by the counterions and coions, respectively. The detailed derivation of membrane potential and energy conversion efficiency with the consideration of Ca$^{2+}$ ions is provided in the supporting information.

**3. Results and Discussion**

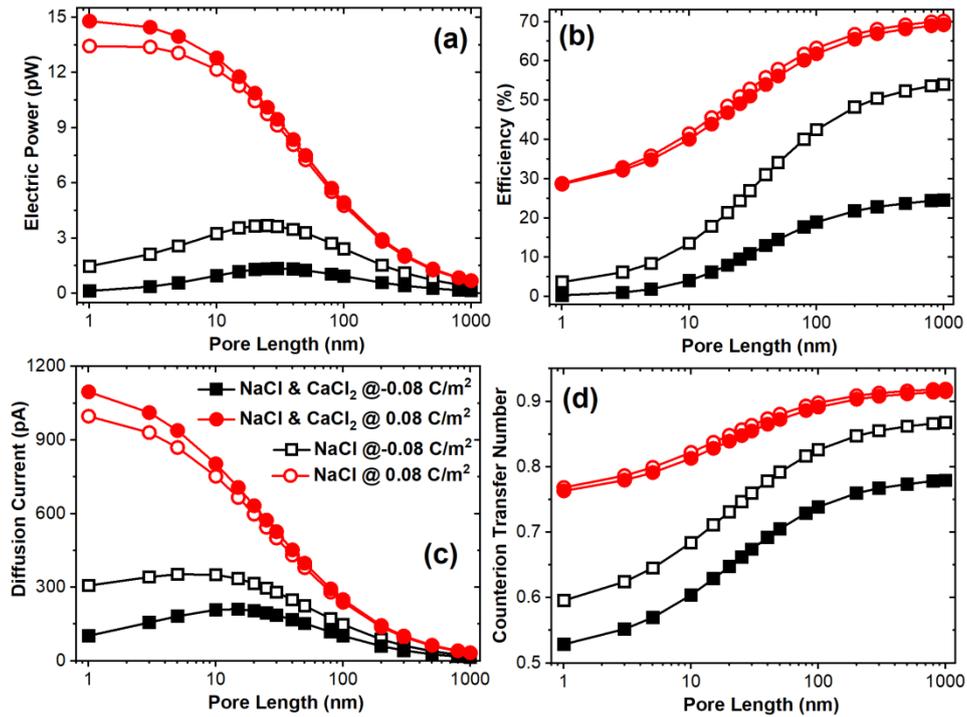

Figure 2. Simulated performance of osmotic energy conversion in negatively and positively charged nanopores with various pore lengths. Cases with and without CaCl$_2$ are considered. (a) Electric power ($P_{max}$), (b) OEC efficiency ($\eta$), (c) diffusion current ($I$), and (d) counterion transfer number. The pore diameter is 10 nm. For mixed solutions, the concentration of CaCl$_2$ is 50:1 mM. The diffusion current contributed by counterions and coions were shown in Figure S5.

Figure 2 shows the simulated OEC performance obtained from nanopores with various lengths under different simulated conditions. Both negatively and positively charged nanopores have been considered, of which all the pore walls are charged. For the cases with a negatively charged nanopore under 500:10 mM NaCl solutions, Na$^+$ ions act as counterions and are the main current carriers. With the pore length varying from 1 to 1000 nm, the output electric power has an increase-decrease profile with a peak value at $L$~30 nm, which results from the increased transfer number of counterions, i.e. ionic selectivity, and the diffusion current decreases as pore length increases.[10] Due to the positive correlation of efficiency to the ionic selectivity of nanopores, the OEC efficiency shares the same trend as that of the transfer number. As shown by our earlier work,[10] for the nanopores with sub-100 nm, the charged exterior membrane surface on the low-concentration side plays an important role during the osmotic energy conversion, which can increase the ionic permeability and selectivity of the nanopore simultaneously.

While in positively charged nanopores the electric power and output efficiency decreases and increases with the pore length, respectively. The OEC performance gets promoted remarkably which is attributed to the increased diffusion current and transfer number due to the larger diffusion coefficient of $Cl^-$ ions. Compared to the negatively charged nanopore, the maximum power and OEC efficiency of positively charged nanopores are increased by ~1.5 and ~1.0 times with 30 nm in length, respectively. As the main carrier ions, $Cl^-$ ions dominate the entire diffusion process (Figure S5), which maintains the ultra-high ionic flux and induces a high transfer number of counterions.

By adding $CaCl_2$ into the solutions on both sides of the nanopore, the effect of $Ca^{2+}$ ions on the OEC performance has been investigated. From Figures 2a and 2b, when the surface charge density is −0.08 C/m$^2$, $Ca^{2+}$ ions act as counterions besides $Na^+$ ions which lower the OEC performance significantly in both electric power and efficiency. This trend agrees well with the experimental results by Zhang et al.[51] While both profiles with and without $Ca^{2+}$ ions share a similar trend. With 50:1 mM $CaCl_2$ in the solutions, the maximum electric power decreases from ~3.6 to ~1.4 pW, and the corresponding conversion efficiency drops from ~27.0% to ~10.8%. The lower-performance OEC with $Ca^{2+}$ ions is correlated to the suppressed diffusion current through the nanopore and the weak ionic selectivity to counterions of the nanopore, which results from much more penetrated $Cl^-$ ions through the nanopore due to the strong screening effect of $Ca^{2+}$ ions to surface charges. Compared to the case without $Ca^{2+}$ ions, the maximum diffusion current in the case with $Ca^{2+}$ ions decreases from 352 pA to 210 pA, and the transfer number of counterions decreases by ~12% under various pore lengths. When nanopore walls carry positive charges, extra $Ca^{2+}$ ions don't have obvious influences on the OEC performance. In this case, $Ca^{2+}$ ions serve as coions which are not the main diffusion current carriers.

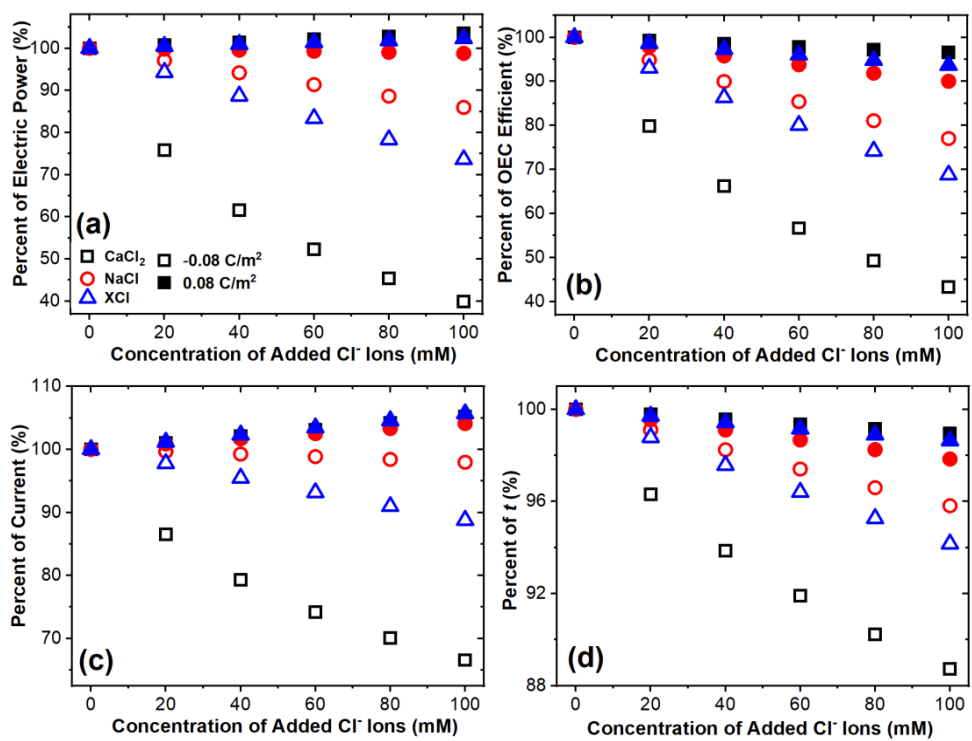

Figure 3. Simulated relative OEC performance in negatively and positively charged nanopores with different extra salts. Extra $CaCl_2$, NaCl, and XCl were added in 500:10 mM NaCl solutions, of which $X^+$ ions are monovalent but with the same diffusion coefficient as $Ca^{2+}$ ions. (a) Electric power, (b) OEC efficiency, (c) diffusion current, and (d) transfer number of counterions. The nanopore diameter and length are 10 nm and 30 nm, respectively. The values of electric power, OEC efficiency, diffusion current, and transfer number were normalized over those from the cases with only 500:10 mM NaCl solutions. Open and solid symbols denote data obtained from negatively and positively charged nanopores, respectively. Actual nanofluidic OEC performance is shown in Figure S6.

From simulation results, the presence of trace $Ca^{2+}$ ions in the NaCl solutions can decrease the OEC performance seriously. This may be due to the increased salt concentration,[50] as well as the lower diffusion coefficient[18] and bivalence of $Ca^{2+}$ ions.[8, 24] A series of simulations were conducted to uncover the underline physical details of the inhibition to the OEC performance from trace $Ca^{2+}$ ions. Different salts were added in 500:10 mM NaCl solutions, i.e. $CaCl_2$, NaCl, and XCl of which $X^+$ ions have the same diffusion coefficient as $Ca^{2+}$ ions. For added NaCl and XCl, the concentration on the high-concentration side varies from 0 to 100 mM, while that for $CaCl_2$ changes from 0 to 50 mM. The concentration ratio of the added salt in the high- and low-concentration reservoirs was maintained at 50. The nanopore dimension was chosen as 10 nm in diameter and 30 nm in length, which exhibited the largest output electric power under 500:10 mM NaCl solutions.

As shown in Figures 3a and 3b, in negatively charged nanopores, the OEC performance decreases in varying degrees with the concentration of the three individual added salts increasing. Compared to the case with 500:10 mM NaCl solutions, when the concentration of added $Cl^-$ ions increases by 100 mM, the electric power and conversion efficiency from the three cases with trace ions decrease by ~14%, 27%, and 60%, as well as by 23%, 32%, and 57%, respectively. From Figures 3c and 3d, the added portion of NaCl lowers the diffusion current and transfer number by ~2% and ~4%, which are caused by the thinner EDLs at higher ionic strength. Though the added XCl is at the same concentration as NaCl, it causes a further reduction in the diffusion current and ionic selectivity by ~10% and ~2% than with added NaCl, respectively. This is mainly due to the lower diffusion coefficient of $X^+$ ions than $Na^+$ ions. For the cases with trace $Ca^{2+}$ ions, besides the above two mechanisms, the bivalence of $Ca^{2+}$ ions can cause another decrease in the diffusion current and transfer number by ~22% and ~5%. We can find that the decreases in the OEC performance caused by trace $Ca^{2+}$ ions are mainly due to their bivalence.

As counterions, divalent $Ca^{2+}$ ions are preferentially attracted to the negatively charged surfaces of the nanopore to form EDLs under electrostatic interactions. The surface charges can be screened much better which induces much more $Cl^-$ ions diffusing across the nanopore from the high-concentration side to the low-concentration side (Figure S6f). Please note that as shown in Figure S7, the concentration of $Ca^{2+}$ ions at 0.5 nm away from the charged wall almost stays constant along the nanopore. This could imply that inside ultra-short nanopores the inhomogeneous charge on the nanopore surface induced by the $Ca^{2+}$ ions can be neglected.[52] Combined with the reduced diffusion flux from $Na^+$ ions, the percent of the counterion transfer number drops from 100% to 89%. While, when the pore surfaces carry positive charges, the OEC performance does not exhibit obvious variation with the added trace salts. This is due to that in all three cases $Cl^-$ ions act as the counterions which are the main diffusion current carriers. The cations with different diffusion coefficients and valence have negligible influences on the diffusion of $Cl^-$ ions.

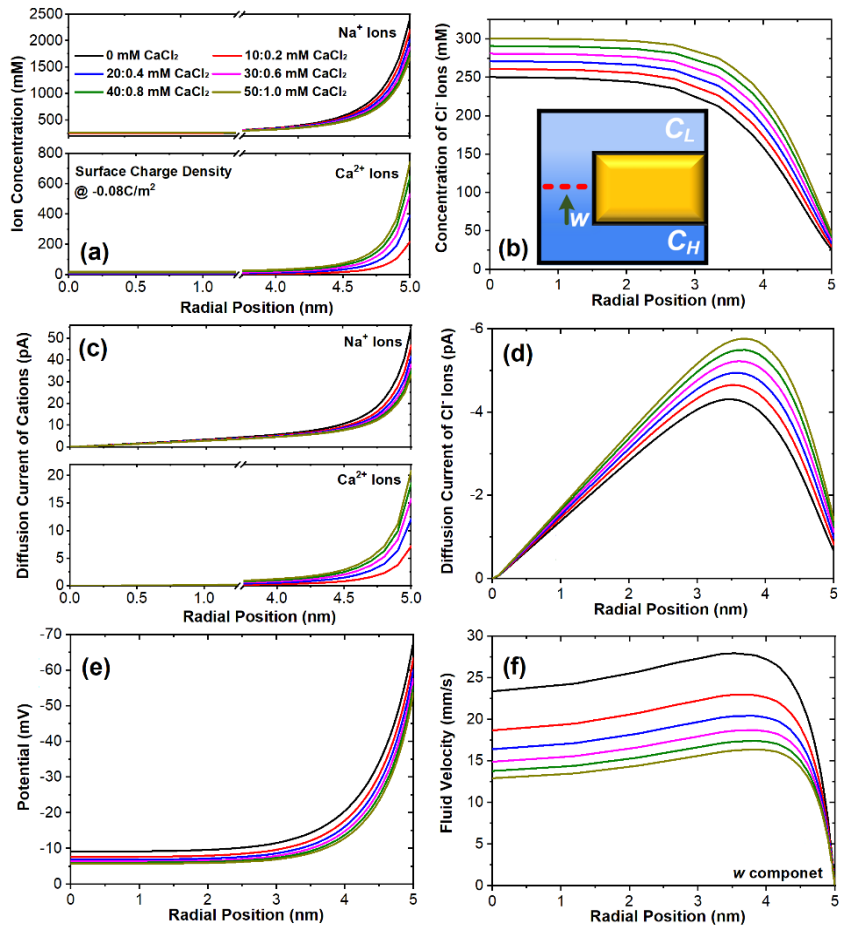

Figure 4. Distributions of ion concentration, diffusion current, potential, and fluid flow inside the nanopore with different concentrations of $CaCl_2$. All data were obtained in the center cross-section of the nanopores. (a-b) Distributions of ion concentration: (a) $Na^+$ and $Ca^{2+}$ ions, and (b) $Cl^-$ ions. The red line in the inset shows locations where the results were obtained. (c-d) Distributions of diffusion current: (c) $Na^+$ and $Ca^{2+}$ ions, and (d) $Cl^-$ ions. (e) Distributions of potential, and (f) fluid velocity i.e. axial component $w$ of fluid flow. The nanopore diameter and length are 10 nm and 30 nm, respectively.

Figure 4 shows the detailed ionic behaviors inside the nanopore during the simulated OEC process with added $CaCl_2$. We have investigated the distributions of ion concentration and flux, as well as electric potential and fluid velocity at the center cross-section of the nanopore. For negatively charged nanopores, with the concentration of $CaCl_2$ increasing from 10 to 50 mM, much more $Ca^{2+}$ ions accumulate in the EDLs exhibiting a sharp increase in the concentration within ~0.5 nm away from the pore wall. This is mainly due to the stronger electrostatic interaction between surface charges and $Ca^{2+}$ ions. Because of the electrical neutrality in the EDLs, the extra $Ca^{2+}$ ions induce a decrease in the concentration of $Na^+$ ions by ~31.3% at the pore wall. As shown in Figure S8a, within 0.5 nm beyond the pore wall, $Ca^{2+}$ ions take more percentages in the overall ion concentration than that in the bulk solutions. As the bulk concentration of $CaCl_2$ increases, the ratio of the $Ca^{2+}$ ion concentration at the charged surface to the total counterion concentration increases from ~16% to 47%. For $Cl^-$ ions, the concentration in the center of the nanopore increases from ~250 to 300 mM with the added 50 mM $CaCl_2$. In the EDLs region, due to electrostatic repulsion $Cl^-$ ions represent a much lower concentration.

The ion current distribution is obtained by integrating the ionic flux over each ring region with a width of 0.1 nm in the center cross-section of the nanopore (Figures 4c and 4d). Similar to our earlier work,[10, 46] EDLs regions near charged surfaces provide a fast passageway for the diffusion of counterions. With the concentration increasing of $Ca^{2+}$ ions, the diffusion current of $Na^+$ ions decreases slightly, while the contribution from $Ca^{2+}$ ions to the total diffusion current of counterions increases. Due to the lower diffusion coefficient, $Ca^{2+}$ ions take a relatively lower percentage in the total diffusion current than in the total counterion concentration. With 50:1 mM $CaCl_2$ added in 500:10 mM NaCl, the contribution from $Ca^{2+}$ ions near the inner-pore surface accounts for ~38% of the total local diffusion current. While the diffusion transport of $Cl^-$ ions as coions mainly happened in the region ~0.5 to 3 nm beyond the pore walls. With the concentration of $CaCl_2$ increasing to 50 mM, the highest local diffusion current of $Cl^-$ ions increases by ~34% from 4.3 to 5.8 pA, which is much higher than the increase in concentration. Due to the stronger electrostatic screening of $Ca^{2+}$ ions to the surface charges (Figure 4e), more $Cl^-$ ions approach the surface which diffuses across the pore under the diffusion-osmotic flow induced by counterions.[53] As shown in Figure 4f, the fluid velocity in the center cross-section of the nanopores decreases as the $Ca^{2+}$ ion concentration increases. Compared to the case with only NaCl, the maximum axial fluid flow is inhibited by 41%. The position with the maximum fluid velocity gradually shifts to the nanopore wall. Due to the significant accumulation of $Ca^{2+}$ ions, the EDLs become thinner, and the diffusion transport of $Na^+$ ions in EDLs is weakened.

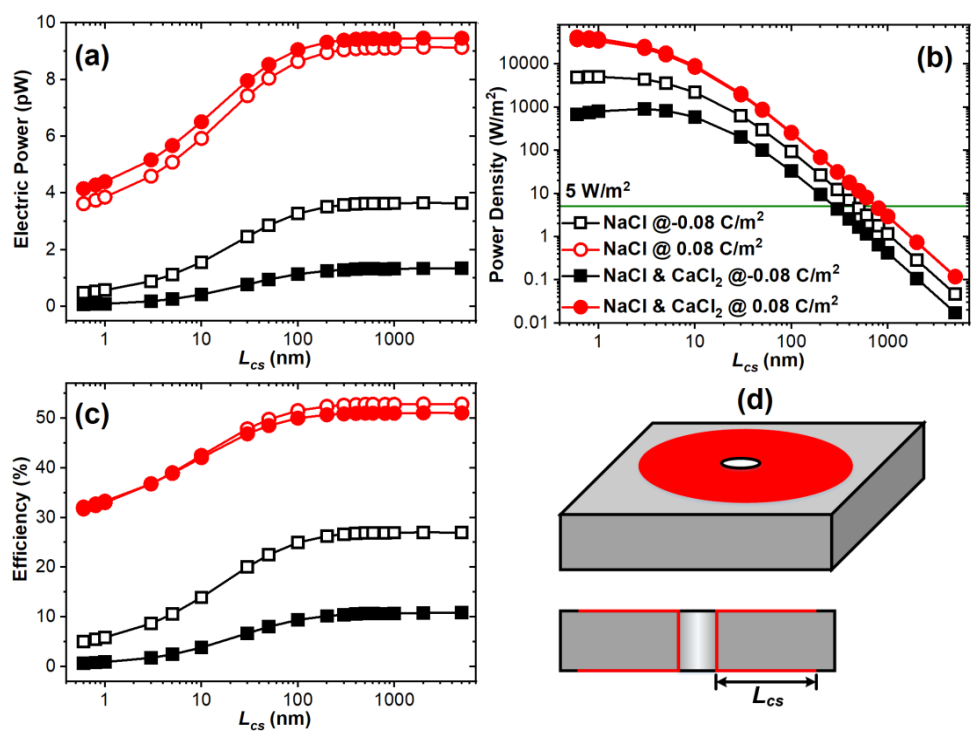

Figure 5. Simulated OEC performance and characteristics of ionic transport in nanopores with negatively or positively charged inner surface and exterior membrane surfaces of various widths ($L_{cs}$). Cases with and without $CaCl_2$ were considered. (a) Electric power, (b) electric power density calculated through $P_{max}/[\pi(R+L_{cs})^2]$, and (c) OEC efficiency. (d) Simulation scheme. Charged surfaces are shown in red. Pore diameter and length were 10 and 30 nm. 50:1 mM $CaCl_2$ were added to 500:10 mM NaCl solutions, respectively.

Due to the enhancement of charged exterior membrane surfaces to the ion diffusion across the nanopores,[10, 46, 47] the effective charged area near pore boundaries is of great importance which determines the tradeoff between porosity and the improvement in OEC performance of single nanopores. The investigation of influences from the width of charged ring regions ($L_{cs}$) beyond nanopores on OEC performance was conducted (Figure 5d), which we think can provide a useful guide to the design of porous membranes for practical applications.[54]

As shown in Figure 5, both the electric power and conversion efficiency are improved pronouncedly with the increase of $L_{cs}$ and approach plateaus at $L_{cs}$~200 nm, which is ~95% of the OEC performance with fully charged exterior nanopore membranes. The power density is roughly evaluated with $P_{max}/[\pi(R+L_{cs})^2]$. As shown in Figure 5b, from negatively charged nanopores, the obtained power density almost shares a constant value with $L_{cs}$ less than 5 nm, which exhibits a decreasing trend as $L_{cs}$ increases further. At $L_{cs}$ ~450 nm, the power density reaches the commercial benchmark of 5 W/m$^2$. In the cases with 50:1 mM CaCl$_2$, the appearance of Ca$^{2+}$ ions inhibit the OEC performance significantly which induces a power density of 5 W/m$^2$ at $L_{cs}$ ~280 nm, which means a higher nanopore density is preferred for the cases with Ca$^{2+}$ ions. Due to the relatively larger diffusion coefficient of Cl$^-$ ions, the OEC performance from positively charged nanopores is much better than that in negatively charged ones. The power density decreases monotonously with $L_{cs}$, which reach the commercial benchmark at $L_{cs}$ ~750 nm. In these cases, the presence of divalent cations has negligible influences on the OEC performance with porous membranes.

## 4. Conclusions

Systematic finite element simulations have been conducted to investigate the detailed influences of the trace divalent ions in seawater/river water on the osmotic energy conversion. In both negatively and positively charged nanopores, divalent Ca$^{2+}$ ions were considered as counterions and coions. When the nanopore is negatively charged, added 50:1 mM Ca$^{2+}$ ions inhibit the output electric power and OEC efficiency significantly. This is mainly due to the bivalence of Ca$^{2+}$ ions besides their lower diffusion coefficient than Na$^+$ ions, instead of the uphill ionic transport.[26] Compared to Na$^+$ ions, Ca$^{2+}$ ions are attracted by negatively charged walls preferentially due to the stronger electrostatic interaction between them and surface charges. Though the bulk concentration of Ca$^{2+}$ ions is as low as 2% to 10% of that of Na$^+$ ions, the accumulation of Ca$^{2+}$ ions in EDLs can take ~16% to 47% in the total amount of counterions, which shields the surface charges effectively. More obvious diffusion of Cl$^-$ ions is induced, which decreases the total diffusion current and ionic selectivity of the nanopore simultaneously. However, in positively charged nanopores, Ca$^{2+}$ ions as coions do not affect the OEC performance. Finally, the promotion of charged exterior membrane surfaces on the OEC has been investigated for ultra-short nanopores. The effective charged region on the membrane surfaces is found to be ~200 nm in width beyond the pore boundaries which is independent of the presence of Ca$^{2+}$ ions in the solutions. Our results present physical details for understanding of effects from divalent ions on OEC. For practical osmotic energy harvesting, multivalent counterions should be filtered.

**Supporting Information**

Simulation details and model verification, methods to obtain membrane potential and diffusion current, more supported simulation results of ionic behaviors in nanopores, and the performance of osmotic energy conversion.

**Acknowledgments**

The authors thank the support from the National Natural Science Foundation of China (Grant No. 52105579), the Natural Science Foundation of Shandong Province (ZR2020QE188), the Natural Science Foundation of Jiangsu Province (BK20200234), the Guangdong Basic and Applied Basic Research Foundation (2019A1515110478), the Qilu Talented Young Scholar Program of Shandong University, Key Laboratory of High-efficiency and Clean Mechanical Manufacture at Shandong University, Ministry of Education, and the Open Foundation of Key

Laboratory of Ocean Energy Utilization and Energy Conservation of Ministry of Education (Grant No. LOEC-202109).

**References**

1. Siria, A.; Bocquet, M.-L.; Bocquet, L. New Avenues for the Large-Scale Harvesting of Blue Energy. *Nat. Rev. Chem.* **2017**, *1*, 0091.
2. Xiao, K.; Jiang, L.; Antonietti, M. Ion Transport in Nanofluidic Devices for Energy Harvesting. *Joule* **2019**, *3*, 2364-2380.
3. Kim, D.-K.; Duan, C.; Chen, Y.-F.; Majumdar, A. Power Generation from Concentration Gradient by Reverse Electrodialysis in Ion-Selective Nanochannels. *Microfluid. Nanofluid.* **2010**, *9*, 1215-1224.
4. Guo, W.; Cao, L.; Xia, J.; Nie, F. Q.; Ma, W.; Xue, J.; Song, Y.; Zhu, D.; Wang, Y.; Jiang, L. Energy Harvesting with Single‐Ion‐Selective Nanopores: A Concentration‐Gradient‐Driven Nanofluidic Power Source. *Adv. Funct. Mater.* **2010**, *20*, 1339-1344.
5. Brauns, E. Salinity Gradient Power by Reverse Electrodialysis: Effect of Model Parameters on Electrical Power Output. *Desalination* **2009**, *237*, 378-391.
6. Dartoomi, H.; Khatibi, M.; Ashrafizadeh, S. N. Nanofluidic Membranes to Address the Challenges of Salinity Gradient Energy Harvesting: Roles of Nanochannel Geometry and Bipolar Soft Layer. *Langmuir* **2022**, *38*, 10313-10330.
7. Vlassiouk, I.; Smirnov, S.; Siwy, Z. Ionic Selectivity of Single Nanochannels. *Nano Lett.* **2008**, *8*, 1978-1985.
8. Israelachvili, J. N., *Intermolecular and Surface Forces*. 3rd ed.; Academic Press: Burlington, MA, 2011.
9. Fan, B.; Bandaru, P. R. Possibility of Obtaining Two Orders of Magnitude Larger Electrokinetic Streaming Potentials, through Liquid Infiltrated Surfaces. *Langmuir* **2020**, *36*, 10238-10243.
10. Ma, L.; Li, Z.; Yuan, Z.; Wang, H.; Huang, C.; Qiu, Y. High-Performance Nanofluidic Osmotic Power Generation Enabled by Exterior Surface Charges under the Natural Salt Gradient. *J. Power Sources* **2021**, *492*, 229637.
11. Zhang, Z.; Wen, L.; Jiang, L. Nanofluidics for Osmotic Energy Conversion. *Nat. Rev. Mater.* **2021**, *6*, 622-639.
12. Long, R.; Zhao, Y.; Kuang, Z.; Liu, Z.; Liu, W. Hydrodynamic Slip Enhanced Nanofluidic Reverse Electrodialysis for Salinity Gradient Energy Harvesting. *Desalination* **2020**, *477*, 114263.
13. Kwon, K.; Lee, S. J.; Li, L.; Han, C.; Kim, D. Energy Harvesting System Using Reverse Electrodialysis with Nanoporous Polycarbonate Track-Etch Membranes. *Int. J. Energy Res.* **2014**, *38*, 530-537.
14. Ji, J.; Kang, Q.; Zhou, Y.; Feng, Y.; Chen, X.; Yuan, J.; Guo, W.; Wei, Y.; Jiang, L. Osmotic Power Generation with Positively and Negatively Charged 2d Nanofluidic Membrane Pairs. *Adv. Funct. Mater.* **2017**, *27*, 1603623.

**TOC**

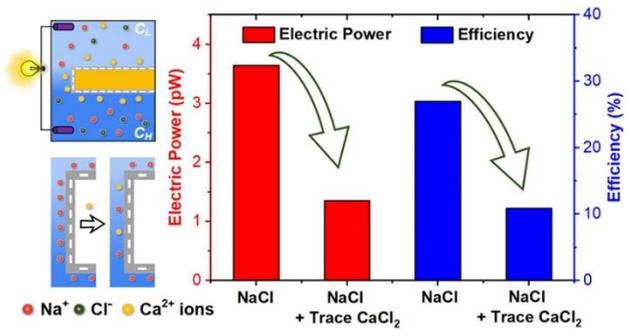